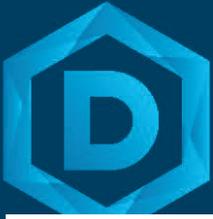

# HELL DIVERS

## The Dark Future of Next-Gen Asymmetric Warfighting

Adam Dorian Wong
adam.wong[at]trojans.dsu.edu
(@MalwareMorghulis)

v1.0

Date: 12 MAR 2024

# Introduction

What if we wake up one day and realize we no longer have dominance over the Five Domains of Warfare (*Sea, Air, Land, Space, Cyber*)? That day may be rapidly approaching. The US is facing numerous international crises that place great strain upon our global security commitments: mass migration, cartel fiefdoms, border crises in Latin America, attacks on the global economy by Houthi proxies, Iranian drone strikes, dominos of coups and resurgence of violent extremist organizations (VEOs) in Africa, continued Russian aggression in Europe, nuclear threats by North Korea, and maritime hostilities by China. The natural order of the world is being threatened. What if one day our enemies – whom we've isolated and denigrated, suddenly band together to challenge the Post-WWII Western-led hegemony? Collectively, these adversaries could potentially meet parity in combat power when juxtaposed with American military might. America has been responsible for global security and stability. Facing involvement in numerous conflicts and fledging recruiting & morale crises, technology and strategy must evolve for us to continue meeting these security commitments.

# AI & Drone Warfare

***Weaponization of Disposable COTS***. Commercial-Off-the-Shelf (COTS) drones have been retrofitted to conduct ad-hoc Intelligence, Surveillance, & Reconnaissance (ISR) missions, retrofitted to drop overhead munitions, or divebomb hard targets [1] [2]. Some have been used to sink warships of greater tonnage and value [3]. Even severely wounding servicemembers can render entire units combat ineffective. The US Military must be cognizant of sourcing of COTS due to possible tampering or competing goals [4] [5]. Swarm drones may eventually be used for mass networking, blanket ISR, or as cumulative direct attack munitions to engage larger well-defended targets [6].

***AI Augmentation of Forces***. Artificial Intelligence (AI) will complement these intersecting areas and domains of war. Autonomous weapons platforms, tactical kits, and big data analyses will provide cutting edge capabilities to warfighters, especially in an era of low recruitment. Our greatest weapons will always be our people. Our greatest priority must always be training and welfare of our service members. Although, we face a recruiting crisis, policy must change to envision augmenting entire formations with drones to compensate for lack of manpower [7]. AI will be a force multiplier from drone-wingman, target acquisitions rifle scopes, "big data" analyses, to acting as a virtual Opposing Forces (OPFOR) for wargaming [8] [9] [10]. The US must be wary of AI hallucinations, confusion, or bias when allowing AI to analyze data, dictate Military Decision-Making Processes (MDMP), or autonomously operate lethal munitions, and the ethical implications of the same.

***Asymmetrical Investment***. Work smarter, not harder. Drones are cheap. Aircraft and large tonnage warships of the *blue-water* navy are expensive. As seen in Ukraine, surface and subsurface sea drones will easily defeat ships at fractions of cost [11]. The future resides in smaller ships that can penetrate enemy waters and provide either high firepower or disable enemy warships. Drones will be necessary for swarming adversarial forces, disabling vehicles, or conducting ISR activities.

*Counter-Drone*. Ultimately, drones add a new level of lethality to the battlespace. Smart-scopes and jamming equipment will be necessary to survive against enemy drone capabilities [12]. Conversely, the US Military needs a method of IFF deconfliction to prevent accidentally downing its own assets from adjacent units [13]. The Joint Counter-UAS School (JCS) must collaborate with Special Operations and allied partners abroad to understand novel implementations of threats [14].

*Cottage Industry*. Additionally, unconventional warfare tactics: hotwiring vehicles, retrofitting drones, or repurposing munitions must lead towards both joint or component-specific agile skunkworks labs and formally training the conventional forces to operate in austere conditions or contested environments and limited support (similar to Ranger School), in accordance with SOF Truths [15]. Tactical lessons will spur agility and innovation that industry simply cannot match [16].

## The Four Digital Horseman

The *Four Digital Horseman* encompass logical and cyber-physical systems: *cyber*, *information*, *electronics*, and *space* – all of which can integrate with AI. Emerson Brooking suggests cyber operations are typically rapid in nature to achieve strategic intelligence advantages whereas information operations shape and influence public narratives longitudinally [17]. Space will dictate dominance over global visibility and communication. Electronic warfare will determine accuracy of guided-munitions and tactical surveillance advantages.

*Cyber Operations (CO)*. Cyber is an enabler of ISR and disrupter of the battlespace. It encompasses networks, servers, hosts, and phones. Cellphones and social media apps will impose OPSEC threats to troops [18] [19]. Lessons from successive iterations of Jack Voltaic have taught us that industrial control systems (ICS) critical infrastructure may inhibit the ability to project force and deploy troops, if they have to worry about activities at home [20] [21]. Adversaries will continue to use cyberespionage to acquire technological or combat advantages [22] [23].

*Information Operations (IO)*. Narratives will be driven by adversaries sewing confusion through social media. As we recoil from two decades of sustained combat in USCENTCOM, we must work hard to restore trust with the American People. The reasons we go to war matter. First, we always stay on the moral high-ground in conflicts and be prepared for careful declassification of information to prove: *jus ad bellum*. As Matt Damon says in *Green Zone*: "*what happens when we need people to trust us again?*" Second, news and social media platforms have fallen under golden mean or false balance fallacies and allowed perpetuation of bad narratives: where VEOs are praised instead of rebuked [24] [25] [26] [27]. *What happens when adversaries play victim because we are creatures of emotion rather than logic*? Third, some bad actors even contribute to adversarial misinformation campaigns as *useful idiots* to beguile the American people [28]. Fourth, false narratives degrade public trust, damage imaging, and sew strive or discord such as with the 2016 "Lisa Case" [29]. The US must correct the narrative without inhibiting transparency and open knowledge. Industry must openly share, identify, and deconstruct infrastructure of mal-influence campaigns with the Federal Partners (DoD, DoJ, DHS).

***Electronic Warfare (EW)***. Initiative is the ability to impose our will upon the enemy forces. Jamming will be a critical issue in disrupting munitions or equipment that leverages wireless communication (satellites, Bluetooth, cellular, Wi-Fi, etc.) [30]. The US and NATO must be prepared to overcome jamming attacks and frequency acquisitions [31] [32]. Additionally, quantum computing will place reliance of cryptographic keys into question.

***Space Warfare***. Reliance on technology will cripple the ability to communicate or navigate. Technology is vulnerable to electromagnetic pulse (EMP) from threats either natural (solar radiation) or artificial (adversarial action) [33]. Satellites could be directed out of orbit by ground controllers such as when they are decommissioned. Orbiting space debris has the potential to damage satellites. The problem is time – time needed to replace and re-deploy space assets. Networks and communications must be redundant across all levels: underground, undersea, wireless, fiber, aerial, near-space, and space.

## Survivability in the Kinetic Battlespace

Fluidity of initiative within combat engagements falls along a spectrum between positional (static) and maneuver (dynamic) warfare. Our survival predisposes need for rapid displacement to avoid incoming attacks and respond in kind.

***Breaking Positional Warfare***. When maneuver warfare decelerates to a slugging match through entrenched forces, the war becomes positional. Ukrainian General Valerii Zaluzhny suggests technological advantages breaks static nature of positional warfare [34]. At parity, it is worth considering whether holding terrain is worth the cost of lives. During WWII, Admiral Chester Nimitz, General Douglas MacArthur, and Admiral Bill Halsey leveraged *island-hopping* to bypass fortified Japanese positions [35]. This idea can be reused in land-based warfare and in USINDOPACOM. This predicates friendly forces can successfully evade enemy electronic warfare and indirect fires. The goal would be to evacuate civilians while enticing to and besieging enemy forces within cities [36]. Friendly forces must leverage strategic flanking to avoid being fixed: either through seaborne or airborne insertions. General MacArthur's landing at Inchon during Operation CHROMITE or General Schwartzkopf's attack during Operation DESERT STORM [37] [38]. Engagements with conventional peer adversaries must not result in eternal gridlocks.

***Maneuver Warfare***. The Principles of War are known by MOOSEMUSS – *Mass*, *Objective*, *Offensive*, *Surprise*, *Economy of Force*, *Maneuver*, *Unity of Command*, *Security*, and *Simplicity* [39]. If only one can be identified, then maneuver will be instrumental in surviving the next war. Precisions Fires and mobility (ability to displace quickly) will dominate battlefields. Systems like BAE's Archer Howitzer System and Lockheed Martin's HIMARS will be key in knocking out Enemy C2 and supply depots. Fires need to be up-armored and lighter to traverse undesirable terrain [40]. Traditional doctrine states: "*fire without maneuver is indecisive*" and "*maneuver without fire is fatal*". Today, indecision will be fatal.

# Future in Supply-Chain Logistics

   *Replacement Theory*. The Defense Industrial Base (DIB) needs to be prepared to convert or expand production to meet scales not seen since WWII. *Just-in-Time* logistics will cripple the ability to sustain combat [41]. The reduction of defense contractors in the 1990s has led to inadequate supplies, productions, and prices [42] [43]. America needs to restore its *Arsenal of Democracy* to replace destroyed equipment or replenish ammunition stockpiles in preparation for Large-Scale Combat Operations (LSCO) on multiple fronts or Multi-Fronted Global Commitments [44] . We must be prepared to not only replace our weapons and ammunition, but that of our allies – in preparation for global conflicts. Ammunition will be expended at unsustainable and exponential rates following initial exchange of strikes [45]. The defense industry is ill-prepared to support long-term high-intensity engagements [46].
   ***Intellectual Property***. The US needs to be cognizant of theft of intellectual property through industrial or cyber espionage. The next great war will be fought against an adversary that achieves or exceeds parity [47]. Adversaries will continue to undermine our kinetic advantages through asymmetrical cyber espionage [48]. The US must be wary of foreign investors and continue to advance research in technologies.
   ***Commitment of Industry***. The military cannot operate without civilian relations with industry. The nation must demonstrate unity on the global stage. Our nation must be vigilant to the commitments of corporations, contractors, and their leaders (some with international presence), which may have conflicting interests and questionable motives – and may interfere with military operations out of ego [49] [50]. Price-gouging is another concern where private industry and defense sector have incessant need for ever-growing or all-time high profit margins [51] [52]. These same DIB contractors have been chided by Admiral Daryl Caudle for failing to meet obligations for weapons productions [53]. Furthermore, some corporations are concerned with money and attempt to maintain a seat of neutrality in conflicts [54]. Companies fall into the false balance fallacy when they are attacked by APT Groups and prohibit leaks of the same hostile threat actors [55]. This could be duplicitous in nature or simply overcorrection of neutrality.
   ***Hardware Contamination***. Securing supply-chains will protect the integrity of the hardware used by servicemembers. Our government must be vigilant of sensitive electronics or counterfeit parts being sourced from less-friendly nations [56]. Conversely, denying adversarial nations our domestic electronic components [57].
   ***Vulnerability of Manufacturing***. Ammunition production plants far from front-lines will become lucrative targets, as in conventional wars like WWII. Production manufacturing accidents are dangerous and lack of diversity in contractors create single points-of-failure [58].

# Future of Domestic Policy

***Migration***. We must be weary of border crises because they may overwhelm social services or provide avenues of infiltration. Migration may be a direct consequence of soft-power projected through influence operations. Simply put, using propaganda to compel mass migration and to overwhelm a target nation [59] [60] [61]. Thereby, to distract or to redirect attention of governing officials to domestic affairs. However, the problem of influx of migration may provide an opportunity for the US to develop new infrastructure and new cities while stimulating the domestic economy.

***Water & Utilities***. America's aquifers are drying up. The US needs to consider potable water as a national asset, glaciers are receding, rivers are changing, and freshwater composes 2.5% of the Earth's surface [62] [63]. Desalination and water purification will be key to maintaining agriculture, livestock, sanitation, and society. Water and energy sectors are equally at risk of cyber-attacks [64] [65] [66].

***Domestic & Stability Operations***. As we traverse geopolitical uncertainty, individual states and territories must have prepositioned caches or stockpiles of equipment to handle CBRN disasters and mass casualty events, in coordination with DHS and FEMA. Additionally, adversaries attempted to subvert the rule of law and apply their controls outside of their domestic borders [67]. Foreign nationals or companies attempted to purchase land near US government installations [68]. Internally, crime may be influenced by poor economic opportunity [69]. Some countries have resorted to strong-man tactics to suppress violent crime and consolidate power [70]. The country must be prepared to maintain order, sustain the economy that benefits the citizenry, and be aware of foreign actors attempting to invest curious locations.

# Predictions of Next-Gen Warfare

***Global Conflict***. We must be prepared to meet our security commitments across the entire Strategic Environment (SE), in a multi-theater war, from which we haven't seen since the Second World War. Saturation attacks through overwhelming firepower or manpower will be prevalent [71] [72]. Some friendly nations may have competing priorities and interests within a given region. At times, these interests align, but when priorities diverge, these associates may become subtly resistant or outright hostile to our nation's ability to respond to geopolitical crises on the associates' doorstep.

***Gray Zone Operations***. Equipment may be repurposed for invasion or create conditions of hybrid warfare. Adversaries will no longer feel obligated to follow Laws of War [73]. Enemies will use re-purposed civilian equipment to smuggle munitions or personnel into place for invasions [74] [75] [76]. Use of proxies will be prevalent to limit attribution or accountability of attacks and prevent encroaching all-out war [77].

***Undersea Data Centers***. Data centers may be positioned underwater to avoid direct attack munitions. It has benefits of using seawater and depth to cool through conduction. Both industry and adversaries have initiated these pilot projects [78] [79].

Replacing assets undersea will be difficult. However, this is not without risk and unknown ecological ramifications by obscuring effects by hiding computers underwater.

   ***Zero-Sum Game***. However, we must be careful with chemically-armed or nuclear-armed adversaries, who may well not adhere to international norms. They may not even care for Mutually-Assured Destruction (MAD). The unfortunate last resort could be "*flipping-the-chessboard*" when confronted with loss of control.

   ***Insurgencies***. When faced with loss, we can expect ill-equipped adversaries to dissolve into guerilla movements. Rendering hostile formations combat ineffective is not enough to stymie loss of initiative. Guerilla uprisings have often crippled conventional forces with weaker, crude, or primitive rudimentary capabilities [80]. Complete destruction of enemy units will prevent prolonged or protracted conflicts. Additionally, we must be cognizant of the intersection of civil-military assets and work to maintain key infrastructure when occupying areas to prevent disorder and insurgencies [81].

   ***Protection of Supply Chains***. Silicon and chips will determine whether we will produce enough smart munitions for prolonged campaigns. The Ukraine War has demonstrated ammunition exhaustion during high-intensity localized conflicts with peer adversaries. These logistics sourcing problems will be exacerbated by multiple conflicts. Civilian industry must understand that conventional war will require mobilization of the society and holistic effort to sustain.

   ***Direct Energy Weapons (DEW)***. Be prepared for non-traditional weapons (direct-energy) harming warfighters or altering the fight [82]. Although comical, smart eye-pro will push tactical information to warfighters and protect from a new threat laser light. Lasers will play a key role in inexpensive C-UAS, C-RAM, and A2AD [83].

   ***Rail Cannons & Mass Drivers***. Magnetically-launched projectiles will be key to reducing overall reliance and expenditure of precious minerals in smart-weapons. Speed and accuracy are necessary to intercept hypersonic missiles without wasting our own.

   ***Decoy Equipment***. Units will be augmented with decoy assets to deceive or confuse enemy ISR [84]. These emplaced shell vehicles will absorb incoming attacks from high-value precision missiles [85]. Enemy missile silos may become shell games to detect or intercept missiles in flight [86] and the US must be judicious on when to intercept in-flight missiles and when to attack pre-fired silos directly.

   ***Smart Cities***. The UN believes two-thirds of the world's population will live in cities by 2030 [87]. Urban warfare will continue to be relevant and highly dangerous for heavier units such as armor. Cities with heavily networked infrastructure pose a danger to maneuver elements through surveillance equipment and AI recognition [88] [89] [90] [91]. It will be difficult to operate in autocratic environments quietly during peacetime.

   ***Tunnel Warfare***. Additionally, the US Military is not prepared to engage enemies in subsurface (or tunnel) warfare. The Vietnam War, the Korean Demilitarized Zone (DMZ), and the recent conflict in Gaza has taught us that subsurface tunnels can still outlast overhead surveillance and provide dangerous avenues of approach [92] [93]. Technology does not nullify threat of primitive capabilities. Tunnels have been used to outflank forces in Avdiivka [94] [95]. Wireless drones will not be able to travel underground due to lack of signal penetration.

   ***Near-Space Balloons***. In 2023, USNORTHCOM shot down a Chinese spy balloon after it traversed the skies above the continental US. Balloons have been used in warfare since the Civil War. Cellular networks need to be secured from misuse. Today, these constitute the threat of *near-space* or the stratus between high-altitude (12mi AGL) to Low Earth Orbit (LEO, 62mi AGL) [96]. Future iterations of warfare may rely upon discreet balloons to maintain redundant networks, in lieu of sending costly or sensitive satellites into orbit [97]. Balloons provide subtle ISR capabilities [98]. A benefit of ballons is that they could be brought back down for maintenance or easier to deploy at-scale. Although vulnerable to missiles, hardware may need to be designed to self-destruct to prevent capture and reverse-engineering. Redundancy and velocity will ensure communications are reestablished.

## Conclusion

Needless to say, a dark future awaits us, but that is no reason to despair. The US must be prepared to engage enemies, whom do not respect international customary law and no longer feel beholden to the status quo of the post-WWII world order. The US military needs to maintain agility in adaptation of new technologies and rethink supporting units. AI and the Four Digital Horsemen will enable or disrupt the ability to respond to crises or within the battlespace. The next great conflict will see a dangerous reliance of electronics and countermeasures of the same. When all smart-weapons are exhausted, fights will turn into fire superiority matches. In any case, redundancy, prepositioned hidden stockpiles, and strong logistics will be paramount to survival. A dark future awaits as the next-generation of warfare will be unpleasant, visceral, a test of ingenuity, and a test of will. As General William T. Sherman once said: "War is Hell", but we will be ready to dive into it.

# References

[1] J. Warrick, "Use of weaponized drones by ISIS spurs terrorism fears," *Washington Post*, May 24, 2023. Accessed: Mar. 10, 2024. [Online]. Available: https://www.washingtonpost.com/world/national-security/use-of-weaponized-drones-by-isis-spurs-terrorism-fears/2017/02/21/9d83d51e-f382-11e6-8d72-263470bf0401_story.html

[2] *Hamas Releases Video of Drone Attack on Israeli Soldiers | News9*, (Oct. 07, 2023). Accessed: Mar. 10, 2024. [Online Video]. Available: https://www.youtube.com/watch?v=dn8XyWxME0E

[3] *CNN reporter goes inside a secret Ukrainian drone workshop*, (Oct. 21, 2022). Accessed: Mar. 10, 2024. [Online Video]. Available: https://www.youtube.com/watch?v=i3by2kNHGgo

[4] E. Baptista, "China's DJI rejects claims of data leaks to Russia on Ukrainian military positions," *Reuters*, Mar. 30, 2022. Accessed: Mar. 10, 2024. [Online]. Available: https://www.reuters.com/world/china/chinas-dji-rejects-claim-that-russian-military-uses-its-drones-ukraine-2022-03-28/

[5] P. Mozur and V. Hopkins, "Ukraine's War of Drones Runs Into an Obstacle: China," *The New York Times*, Sep. 30, 2023. Accessed: Mar. 10, 2024. [Online]. Available: https://www.nytimes.com/2023/09/30/technology/ukraine-russia-war-drones-china.html

[6] M. P. Klymenko Ray Eitel-Porter, and Max, "How Swarm Intelligence Blends Global and Local Insight," MIT Sloan Management Review. Accessed: Mar. 10, 2024. [Online]. Available: https://sloanreview.mit.edu/article/how-swarm-intelligence-blends-global-and-local-insight/

[7] "Chiefs Discuss Military Recruiting Challenges at Committee Hearing," U.S. Department of Defense. Accessed: Mar. 10, 2024. [Online]. Available: https://www.defense.gov/News/News-Stories/Article/Article/3610846/chiefs-discuss-military-recruiting-challenges-at-committee-hearing/https%3A%2F%2Fwww.defense.gov%2FNews%2FNews-Stories%2FArticle%2FArticle%2F3610846%2Fchiefs-discuss-military-recruiting-challenges-at-committee-hearing%2F

[8] T. South, "Army finally picks an optic for Next Generation Squad Weapon," Army Times. Accessed: Mar. 10, 2024. [Online]. Available: https://www.armytimes.com/news/your-army/2022/01/07/army-finally-picks-an-optic-for-next-generation-squad-weapon/

[9] M. Pomerleau, "US Cyber Command advances on platform to consolidate its myriad tools and data," C4ISRNet. Accessed: Mar. 10, 2024. [Online]. Available: https://www.c4isrnet.com/cyber/2020/11/01/us-cyber-command-advances-on-platform-to-consolidate-its-myriad-tools-and-data/

[10] "New in 2024: Air Force plans autonomous flight tests for drone wingmen." Accessed: Mar. 10, 2024. [Online]. Available: https://www.defensenews.com/air/2023/12/30/new-in-2024-air-force-plans-autonomous-flight-tests-for-drone-wingmen/

[11] "Ukraine claims it has sunk another Russian warship in the Black Sea using high-tech sea drones," AP News. Accessed: Mar. 10, 2024. [Online]. Available: https://apnews.com/article/russia-ukraine-war-sea-drones-a316215f5d9fc88a488ef2d99eb828c9

[12] "MOD issues AI technology to help Colchester soldiers shoot down drones," Mar. 06, 2024. Accessed: Mar. 10, 2024. [Online]. Available: https://www.bbc.com/news/uk-england-essex-68484830

[13] *Combat Vets from Israel Explain Urban Warfare, Drones, Insurgency tactics*, (Mar. 10, 2024). Accessed: Mar. 10, 2024. [Online Video]. Available: https://www.youtube.com/watch?v=gdQS97fJY4Q

[14] *Inside the U.S. Military's New Drone Warfare School | WSJ*, (Jan. 02, 2024). Accessed: Mar. 10, 2024. [Online Video]. Available: https://www.youtube.com/watch?v=xb5qMvie9sU
[15] "SOF Truths." Accessed: Mar. 10, 2024. [Online]. Available: https://www.socom.mil/about/sof-truths
[16] G. McCardle, "Ukraine's Cottage Industry of Improvised Weapons, Hedgehogs and Victory Beer," SOFREP. Accessed: Dec. 27, 2023. [Online]. Available: https://sofrep.com/news/ukraines-cottage-industry-of-improvised-weapons-hedgehogs-and-victory-beer/
[17] E. Brooking, "Why Your Information Operation Is Probably a Bad Idea," CYBERWARCON. Accessed: Feb. 02, 2024. [Online]. Available: https://www.cyberwarcon.com/why-your-information-operation-is-probably-a-bad-idea
[18] S. B. Ankel Sophia, "Russian spies are using Tinder to ensnare German soldiers and politicians to get them to disclose Ukraine war secrets, counterintelligence warns," Business Insider. Accessed: Mar. 10, 2024. [Online]. Available: https://www.businessinsider.com/russia-spies-tinder-secrets-from-german-soldiers-politicians-intel-2023-4
[19] A. Yuhas, T. Gibbons-Neff, and Y. Al-Hlou, "For Russian Troops, Cellphone Use Is a Persistent, Lethal Danger," *The New York Times*, Jan. 04, 2023. Accessed: Mar. 10, 2024. [Online]. Available: https://www.nytimes.com/2023/01/04/world/europe/ukraine-russia-cellphones.html
[20] "Jack Voltaic." Accessed: Mar. 10, 2024. [Online]. Available: https://cyber.army.mil/Research/Jack-Voltaic/
[21] Z. Siddiqui, "US disrupts Chinese hacking campaign targeting critical infrastructure, officials say," *Reuters*, Jan. 31, 2024. Accessed: Mar. 10, 2024. [Online]. Available: https://www.reuters.com/technology/us-disrupts-chinese-botnet-targeting-critical-infrastructure-fbi-says-2024-01-31/
[22] "Translation: 14th Five-Year Plan for National Informatization – Dec. 2021," DigiChina. Accessed: Mar. 10, 2024. [Online]. Available: https://digichina.stanford.edu/work/translation-14th-five-year-plan-for-national-informatization-dec-2021/
[23] B. Jensen, "How the Chinese Communist Party Uses Cyber Espionage to Undermine the American Economy," Oct. 2023, Accessed: Mar. 10, 2024. [Online]. Available: https://www.csis.org/analysis/how-chinese-communist-party-uses-cyber-espionage-undermine-american-economy
[24] M. Gault, "Everything We Know About 'Timhouthi Chalamet,' the Yemeni Influencer Celebrating Red Sea Ship Raids," Vice. Accessed: Mar. 10, 2024. [Online]. Available: https://www.vice.com/en/article/epv74k/everything-we-know-about-timhouthi-chalamet-the-yemeni-influencer-celebrating-red-sea-ship-raids
[25] P. W. Singer and E. T. Brooking, "Gaza and the Future of Information Warfare," *Foreign Affairs*, Dec. 05, 2023. Accessed: Mar. 12, 2024. [Online]. Available: https://www.foreignaffairs.com/middle-east/gaza-and-future-information-warfare
[26] "Inside the Israel-Hamas Information War," TIME. Accessed: Mar. 12, 2024. [Online]. Available: https://time.com/6549544/israel-and-hamas-the-media-war/
[27] M. J. Ard, "Hamas Is Winning the Information War." Accessed: Mar. 12, 2024. [Online]. Available: https://www.discoursemagazine.com/p/hamas-is-winning-the-information
[28] J. Rutenberg and M. M. Grynbaum, "Tucker Carlson's Lesson in the Perils of Giving Airtime to an Autocrat," *The New York Times*, Feb. 16, 2024. Accessed: Mar. 12, 2024. [Online]. Available: https://www.nytimes.com/2024/02/16/business/media/tucker-carlson-putin-navalny.html
[29] "NATO Review - The 'Lisa case': Germany as a target of Russian disinformation," NATO Review. Accessed: Mar. 10, 2024. [Online]. Available:

https://www.nato.int/docu/review/articles/2016/07/25/the-lisa-case-germany-as-a-target-of-russian-disinformation/index.html

[30] R. Olearchyk, "Military briefing: Russia has the upper hand in electronic warfare with Ukraine." Accessed: Mar. 10, 2024. [Online]. Available: https://www.ft.com/content/a477d3f1-8c7e-4520-83b0-572ad674c28e

[31] D. Goward and L. Ferran, "As Baltics see spike in GPS jamming, NATO must respond," Breaking Defense. Accessed: Mar. 10, 2024. [Online]. Available: https://breakingdefense.sites.breakingmedia.com/2024/01/as-baltics-see-spike-in-gps-jamming-nato-must-respond/

[32] P. Mozur and A. Krolik, "The Invisible War in Ukraine Being Fought Over Radio Waves," *The New York Times*, Nov. 19, 2023. Accessed: Mar. 10, 2024. [Online]. Available: https://www.nytimes.com/2023/11/19/technology/russia-ukraine-electronic-warfare-drone-signals.html

[33] "The Buzz About Electromagnetic Pulse Weapons," Army Cyber Institute. Accessed: Mar. 10, 2024. [Online]. Available: https://cyber.army.mil/News/Article/3464441/the-buzz-about-electromagnetic-pulse-weapons/https%3A%2F%2Fcyber.army.mil%2FNews%2FArticle%2F3464441%2Fthe-buzz-about-electromagnetic-pulse-weapons%2F

[34] "ZALUZHNYI_FULL_VERSION.pdf." Accessed: Mar. 12, 2024. [Online]. Available: https://infographics.economist.com/2023/ExternalContent/ZALUZHNYI_FULL_VERSION.pdf

[35] "The Pacific Strategy in World War II: Lessons for China's Antiaccess/Area Denial Perimeter," U.S. Naval Institute. Accessed: Mar. 12, 2024. [Online]. Available: https://www.usni.org/magazines/naval-history-magazine/2022/june/pacific-strategy-world-war-ii-lessons-chinas

[36] "What You Need To Know About The Battle Of Stalingrad," Imperial War Museums. Accessed: Mar. 12, 2024. [Online]. Available: https://www.iwm.org.uk/history/what-you-need-to-know-about-the-battle-of-stalingrad

[37] "Inchon Landing (Operation Chromite): September 1950." Accessed: Mar. 12, 2024. [Online]. Available: http://public1.nhhcaws.local/content/history/museums/nmusn/explore/photography/korean-war/inchon-landing.html

[38] "Deception in the Desert: Deceiving Iraq in Operation DESERT STORM." Accessed: Mar. 12, 2024. [Online]. Available: https://www.armyupress.army.mil/Books/Browse-Books/iBooks-and-EPUBs/Deception-in-the-Desert/

[39] F. Kearney and S. Marks, "Leadership and the Principles Of War Applied To Business: Two Sides Of The Same Coin - Thayer Leadership." Accessed: Mar. 10, 2024. [Online]. Available: https://www.thayerleadership.com/blog/2017/leadership-and-the-principles-of-war-applied-to-business

[40] "ARCHER Wheeled Artillery System & Mobile Howitzer," BAE Systems | International. Accessed: Mar. 10, 2024. [Online]. Available: https://www.baesystems.com/en/product/archer

[41] E. Lipton, "From Rockets to Ball Bearings, Pentagon Struggles to Feed War Machine," *The New York Times*, Mar. 24, 2023. Accessed: Jan. 01, 2024. [Online]. Available: https://www.nytimes.com/2023/03/24/us/politics/military-weapons-ukraine-war.html

[42] Chang and M. Chakrabarti, "'The last supper': How a 1993 Pentagon dinner reshaped the defense industry," Mar. 01, 2023. Accessed: Dec. 27, 2023. [Online]. Available: https://www.wbur.org/onpoint/2023/03/01/the-last-supper-how-a-1993-pentagon-dinner-reshaped-the-defense-industry

[43] The Editorial Board, "The Pentagon Has a Supply-Chain Problem," *Bloomberg.com*, Jun. 07, 2022. Accessed: Jan. 01, 2024. [Online]. Available: https://www.bloomberg.com/opinion/articles/2022-06-07/the-pentagon-has-a-supply-chain-problem

[44] "Ukraine's army is suffering artillery 'shell hunger,'" POLITICO. Accessed: Mar. 11, 2024. [Online]. Available: https://www.politico.eu/article/ukrainian-army-suffers-from-artillery-shell-hunger/

[45] G. W. D. W. Editor, "Ret. Admiral sounds alarm about munition shortage as China threat grows," Newsweek. Accessed: Mar. 11, 2024. [Online]. Available: https://www.newsweek.com/ret-navy-admiral-sounds-alarm-about-munition-shortage-china-threat-grows-1850995

[46] S. Skove, "In race to make artillery shells, US, EU see different results," Defense One. Accessed: Jan. 01, 2024. [Online]. Available: https://www.defenseone.com/business/2023/11/race-make-artillery-shells-us-eu-see-different-results/392288/

[47] R. Lucas, "Chinese national arrested and charged with stealing AI trade secrets from Google," *NPR*, Mar. 06, 2024. Accessed: Mar. 11, 2024. [Online]. Available: https://www.npr.org/2024/03/06/1236380984/china-google-fbi-ai

[48] C. Demarest, "US says hackers attacked defense organization, stole sensitive info," C4ISRNet. Accessed: Jan. 01, 2024. [Online]. Available: https://www.c4isrnet.com/cyber/2022/10/05/us-says-hackers-attacked-defense-organization-stole-sensitive-info/

[49] C. Davenport and J. Menn, "Musk refused to allow Ukraine's military to use Starlink to attack Russian fleet," *Washington Post*, Sep. 11, 2023. Accessed: Mar. 12, 2024. [Online]. Available: https://www.washingtonpost.com/technology/2023/09/07/ukraine-starlink-musk-biography/

[50] "Russia is using Starlink in occupied areas, Ukraine says," *Reuters*, Feb. 11, 2024. Accessed: Mar. 12, 2024. [Online]. Available: https://www.reuters.com/world/europe/ukraines-military-intelligence-says-it-confirms-use-musks-starlink-by-russian-2024-02-11/

[51] A. Chasan, "Pentagon falling victim to price gouging by military contractors | 60 Minutes - CBS News." Accessed: Mar. 12, 2024. [Online]. Available: https://www.cbsnews.com/news/pentagon-budget-price-gouging-military-contractors-60-minutes-2023-05-21/

[52] Whitaker, "Weapons contractors hitting Pentagon with inflated prices | 60 Minutes - CBS News." Accessed: Jan. 01, 2024. [Online]. Available: https://www.cbsnews.com/news/weapons-contractors-price-gouging-pentagon-60-minutes-transcript-2023-05-21/

[53] "Navy Secretary Warns: If Defense Industry Can't Boost Production, Arming Both Ukraine and the US May Become 'Challenging,'" Defense One. Accessed: Mar. 12, 2024. [Online]. Available: https://www.defenseone.com/threats/2023/01/navy-secretary-warns-if-defense-industry-cant-boost-production-arming-both-ukraine-and-us-may-become-challenging/381722/

[54] eliasgroll, "Chinese hacking operation puts Microsoft in the crosshairs over security failures," CyberScoop. Accessed: Mar. 12, 2024. [Online]. Available: https://cyberscoop.com/microsoft-china-hacking-state/

[55] ShobanChiddarth, "I-S00N was removed from github, as well as all of it's forks," r/cybersecurity. Accessed: Mar. 12, 2024. [Online]. Available: www.reddit.com/r/cybersecurity/comments/1axsfzv/is00n_was_removed_from_github_as_well_as_all_of/

[56] R. Clements, "Report unveils the presence of more than a million counterfeit electronic parts in U.S. combat planes," The Aviationist. Accessed: Jan. 01, 2024. [Online]. Available: https://theaviationist.com/2012/05/28/counterfeit-parts/

[57] "AMD reportedly hits U.S. regulatory roadblock for Chain-tailored chip." Accessed: Mar. 12, 2024. [Online]. Available: https://www.cnbc.com/2024/03/05/amd-reportedly-hits-us-regulatory-roadblock-for-chain-tailored-chip.html

[58] G. Lubold, "The U.S. Military Relies on One Louisiana Factory. It Blew Up.," *Wall Street Journal*, Apr. 26, 2023. Accessed: Jan. 01, 2024. [Online]. Available: https://www.wsj.com/articles/the-u-s-military-has-an-explosive-problem-6e1a1049

[59] "Russia sending migrants to our border, Estonia says," Nov. 22, 2023. Accessed: Mar. 11, 2024. [Online]. Available: https://www.bbc.com/news/world-europe-67503800

[60] L. Ferré-Sadurní, "Migrants Face Cold, Perilous Crossing From Canada to New York," *The New York Times*, Feb. 11, 2024. Accessed: Mar. 11, 2024. [Online]. Available: https://www.nytimes.com/2024/02/11/nyregion/migrants-canada-northern-border.html

[61] S. Alfonsi, "Chinese migrants are the fastest growing group crossing from Mexico into U.S. at southern border - CBS News." Accessed: Mar. 11, 2024. [Online]. Available: https://www.cbsnews.com/news/chinese-migrants-fastest-growing-group-us-mexico-border-60-minutes-transcript/

[62] M. Rojanasakul, C. Flavelle, B. Migliozzi, and E. Murray, "America Is Using Up Its Groundwater Like There's No Tomorrow," *The New York Times*, Aug. 28, 2023. Accessed: Mar. 11, 2024. [Online]. Available: https://www.nytimes.com/interactive/2023/08/28/climate/groundwater-drying-climate-change.html

[63] "The distribution of water on, in, and above the Earth | U.S. Geological Survey." Accessed: Mar. 11, 2024. [Online]. Available: https://www.usgs.gov/media/images/distribution-water-and-above-earth

[64] "Cyber-Attack Against Ukrainian Critical Infrastructure | CISA." Accessed: Mar. 12, 2024. [Online]. Available: https://www.cisa.gov/news-events/ics-alerts/ir-alert-h-16-056-01

[65] "Exploitation of Unitronics PLCs used in Water and Wastewater Systems | CISA." Accessed: Mar. 12, 2024. [Online]. Available: https://www.cisa.gov/news-events/alerts/2023/11/28/exploitation-unitronics-plcs-used-water-and-wastewater-systems

[66] "CISA, FBI confirm critical infrastructure intrusions by China-linked hackers," Utility Dive. Accessed: Mar. 12, 2024. [Online]. Available: https://www.utilitydive.com/news/cisa-fbi-critical-infrastructure-china-hacker/706979/

[67] M. Rajagopalan and W. K. Rashbaum, "With F.B.I. Search, U.S. Escalates Global Fight Over Chinese Police Outposts," *The New York Times*, Jan. 12, 2023. Accessed: Mar. 12, 2024. [Online]. Available: https://www.nytimes.com/2023/01/12/world/europe/china-outpost-new-york.html

[68] "Foreign purchase of land near U.S. bases would need federal okay under proposed rule," NBC News. Accessed: Mar. 12, 2024. [Online]. Available: https://www.nbcnews.com/politics/national-security/foreign-china-buy-land-us-military-bases-require-government-approval-rcna83152

[69] M. D. Melick, "The Relationship between Crime and Unemployment".

[70] "El Salvador extends anti-gang emergency decree for 24th time. It's now been in effect for two years," AP News. Accessed: Mar. 12, 2024. [Online]. Available: https://apnews.com/article/el-salvador-gang-crackdown-emergency-decree-154d50d40d1f2a46a48b6880624df141

[71] "Intensified Russian airstrikes are stretching Ukraine's air defense resources, officials say," AP News. Accessed: Mar. 11, 2024. [Online]. Available: https://apnews.com/article/russia-ukraine-war-air-defense-missiles-95cf7f351d05c0c4b704d9fe8a0c5f39

[72] P. Dickinson, "Human wave tactics are demoralizing the Russian army in Ukraine," Atlantic Council. Accessed: Mar. 11, 2024. [Online]. Available: https://www.atlanticcouncil.org/blogs/ukrainealert/human-wave-tactics-are-demoralizing-the-russian-army-in-ukraine/

[73] "Watch Out for Little Green Men," Brookings. Accessed: Mar. 10, 2024. [Online]. Available: https://www.brookings.edu/articles/watch-out-for-little-green-men/
Adam Dorian Wong                    -12-
@MalwareMorghulis

[74] "Mind the Gap: How China's Civilian Shipping Could Enable a Taiwan Invasion," War on the Rocks. Accessed: Mar. 10, 2024. [Online]. Available: https://warontherocks.com/2021/08/mind-the-gap-how-chinas-civilian-shipping-could-enable-a-taiwan-invasion/

[75] M. Loh, "China can't use civilian ships to invade Taiwan just yet, but it's getting there, says former US intelligence officer," Business Insider. Accessed: Mar. 10, 2024. [Online]. Available: https://www.businessinsider.com/china-growing-capacity-invade-taiwan-civilian-ships-not-ready-yet-2024-2

[76] J. Goldenziel, "China's Ferry Tale Taiwan Invasion Plan Is A Legal Nightmare," Forbes. Accessed: Mar. 10, 2024. [Online]. Available: https://www.forbes.com/sites/jillgoldenziel/2023/01/31/chinas-ferry-tale-taiwan-invasion-is-a-legal-nightmare/

[77] B. P. and B. Benoit, "WSJ News Exclusive | A New Terror Threat Is Emerging in Europe Linked to Iran, Gaza War," WSJ. Accessed: Mar. 12, 2024. [Online]. Available: https://www.wsj.com/world/europe/a-new-terror-threat-is-emerging-in-europe-linked-to-iran-gaza-war-fb297119

[78] "Microsoft finds underwater datacenters are reliable, practical and use energy sustainably," Source. Accessed: Mar. 12, 2024. [Online]. Available: https://news.microsoft.com/source/features/sustainability/project-natick-underwater-datacenter/

[79] A. Paleja, "China sinks 1400-ton data center in sea with power of 6 million PCs," Interesting Engineering. Accessed: Mar. 12, 2024. [Online]. Available: https://interestingengineering.com/innovation/worlds-first-underwater-data-center-china

[80] "Myanmar's Army is Collapsing: An Update," Council on Foreign Relations. Accessed: Mar. 12, 2024. [Online]. Available: https://www.cfr.org/blog/myanmars-army-collapsing-update

[81] "How Disbanding the Iraqi Army Fueled ISIS," TIME. Accessed: Mar. 10, 2024. [Online]. Available: https://time.com/3900753/isis-iraq-syria-army-united-states-military/

[82] "China Blinded a Coast Guard Crew With a Military-Grade Laser," Popular Mechanics. Accessed: Mar. 10, 2024. [Online]. Available: https://www.popularmechanics.com/military/weapons/a42863850/china-uses-military-grade-laser-blind-philippine-coast-guard/

[83] "Iron Beam: How Israel's New Laser Weapon Works," WSJ. Accessed: Mar. 10, 2024. [Online]. Available: https://www.wsj.com/video/series/news-explainers/iron-beam-how-israels-new-laser-weapon-works/7FEDC20F-B080-4E7E-B3C7-832702D5F191

[84] M. J. Donovan, "Strategic Deception: Operation Fortitude:," Defense Technical Information Center, Fort Belvoir, VA, Jan. 2002. doi: 10.21236/ADA404434.

[85] J. Hudson, "Ukraine lures Russian missiles with decoys of U.S. rocket system," *Washington Post*, Aug. 30, 2022. Accessed: Mar. 10, 2024. [Online]. Available: https://www.washingtonpost.com/world/2022/08/30/ukraine-russia-himars-decoy-artillery/

[86] *This Chinese Military Unit Runs One of the World's Largest Missile Forces | WSJ*, (Jan. 23, 2024). Accessed: Mar. 10, 2024. [Online Video]. Available: https://www.youtube.com/watch?v=m_SgPTgspXQ

[87] "Urbanization." Accessed: Mar. 10, 2024. [Online]. Available: https://www.unfpa.org/urbanization

[88] *How To Escape The City (Urban Evasion While Being Hunted)*, (Oct. 29, 2023). Accessed: Mar. 10, 2024. [Online Video]. Available: https://www.youtube.com/watch?v=_f-0GRismLg

[89] "Third Infiltration Tunnel | The DMZ, South Korea | Attractions," Lonely Planet. Accessed: Mar. 12, 2024. [Online]. Available: https://www.lonelyplanet.com/south-korea/gyeonggi-do/panmunjom-and-the-dmz-tour/attractions/third-infiltration-tunnel/a/poi-sig/1410015/357399

[90] M. Xiao, P. Mozur, I. Qian, and A. Cardia, "Video: China's Surveillance State Is Growing. These Documents Reveal How.," *The New York Times*, Jun. 21, 2022. Accessed: Mar. 10, 2024. [Online].

Available: https://www.nytimes.com/video/world/asia/100000008314175/china-government-surveillance-data.html

[91] P. Mozur, M. Xiao, and J. Liu, "'An Invisible Cage': How China Is Policing the Future," *The New York Times*, Jun. 26, 2022. Accessed: Mar. 10, 2024. [Online]. Available: https://www.nytimes.com/2022/06/25/technology/china-surveillance-police.html

[92] N. Rennolds, "I visited the Cu Chi Tunnels used to ambush US soldiers, a grisly but unforgettable reminder of the Vietnam War — take a look," Business Insider. Accessed: Mar. 10, 2024. [Online]. Available: https://www.businessinsider.com/cu-chi-tunnels-unforgettable-reminder-vietnam-war-viet-cong-2023-6

[93] "Israel finds large tunnel near Gaza border, raises questions about prewar intelligence | AP News." Accessed: Mar. 10, 2024. [Online]. Available: https://apnews.com/article/israel-tunnel-war-gaza-hamas-350dbabc2890ee3c4fc2ec33c2a53f09

[94] *Fall of Avdiivka - Russian Invasion of Ukraine DOCUMENTARY*, (Mar. 12, 2024). Accessed: Mar. 12, 2024. [Online Video]. Available: https://www.youtube.com/watch?v=8f3Ehc-_ILY

[95] "Ukraine: The Battle for Soledar's Salt Mines." Accessed: Mar. 12, 2024. [Online]. Available: https://iwpr.net/global-voices/ukraine-battle-soledars-salt-mines

[96] "Near-Space," Air & Space Forces Magazine. Accessed: Mar. 10, 2024. [Online]. Available: https://www.airandspaceforces.com/article/0705near/

[97] "U.S. intelligence officials determined the Chinese spy balloon used a U.S. internet provider to communicate," NBC News. Accessed: Mar. 10, 2024. [Online]. Available: https://www.nbcnews.com/news/investigations/us-intelligence-officials-determined-chinese-spy-balloon-used-us-inter-rcna131150

[98] A. Holpuch, "A Brief History of Spying With Balloons," *The New York Times*, Feb. 03, 2023. Accessed: Mar. 10, 2024. [Online]. Available: https://www.nytimes.com/2023/02/03/us/spy-balloon-history.html